# Coherence properties of electron beam activated emitters in hexagonal boron nitride under resonant excitation

Jake Horder[1], Simon J. U. White[1], Angus Gale[1], Chi Li[1], Kenji Watanabe[2], Takashi Taniguchi[3], Mehran Kianinia[1,4], Igor Aharonovich[1,4] and Milos Toth[1,4]

1. School of Mathematical and Physical Sciences, University of Technology Sydney, Ultimo, New South Wales 2007, Australia
2. Research Center for Functional Materials, National Institute for Materials Science, Tsukuba 305-0044, Japan
3. International Center for Materials Nanoarchitectonics, National Institute for Materials Science, Tsukuba 305-0044, Japan
4. ARC Centre of Excellence for Transformative Meta-Optical Systems, University of Technology Sydney, Ultimo, New South Wales 2007, Australia

milos.toth@uts.edu.au; igor.aharonovich@uts.edu.au

**Abstract**

Two dimensional materials are becoming increasingly popular as a platform for studies of quantum phenomena and for the production of prototype quantum technologies. Quantum emitters in 2D materials can host two level systems that can act as qubits for quantum information processing. Here, we characterize the behavior of position-controlled quantum emitters in hexagonal boron nitride at cryogenic temperatures. Over two dozen sites, we observe an ultra-narrow distribution of the zero phonon line at ~436 nm, together with strong linearly polarized emission. We employ resonant excitation to characterize the emission lineshape and find spectral diffusion and phonon broadening contribute to linewidths in the range 1-2 GHz. Rabi oscillations are observed at a range of resonant excitation powers, and under 1 µW excitation a coherent superposition is maintained up to 0.90 ns. Our results are promising for future employment of quantum emitters in hBN for scalable quantum technologies.

**Introduction**

The development of quantum technologies is expected to enable opportunities in enhanced sensing, secure communication, and quantum computation[1–4]. Solid-state quantum systems, in particular, have the potential to be efficiently integrated into scalable quantum technology devices[5], and platforms such as diamond color centers have already achieved important milestones in this direction[6,7]. However, interest is growing in the use of layered two dimensional materials such as hexagonal boron nitride (hBN), given their atomically thin nature. The exploration of defects in hBN has shown this platform is capable of hosting both spin[8–13] and single photon emitters (SPEs)[14–21], which could provide the basis for a spin-photon interface for use in quantum information distribution[22].

Despite progress, achieving desirable photonic capabilities from an SPE in hBN continues to impose significant experimental challenges. For example, photon indistinguishability requires an emitter with a spectral linewidth free of inhomogeneous broadening, but finding such a candidate has been non-trivial[23]. Moreover, hBN hosts numerous SPEs with a broad range of emission wavelengths, and most defect fabrication methods are ineffective at generating SPEs consistently with a small variance in emission wavelength. An exception to this problem is a recently-developed electron beam irradiation technique for the generation of so-called blue emitters with an emission wavelength at ~ 436 nm[24–26]. The technique is site-specific, it works reliably with high quality hBN crystals and the emitters can be produced en masse – all of which make the blue emitters appealing for studies of SPE spectral behavior and coherence properties.

In this work, we characterize the non-resonant photoluminescence properties of the blue emitter at cryogenic temperatures. Resonant excitation is then used to characterize the emission lineshape, the degree of spectral diffusion, and linewidth broadening versus temperature. Most significantly, we demonstrate coherence of optical transitions through the observation of Rabi oscillations. This result is a key requisite capability for these blue emitters to be considered as a viable platform for the development of scalable quantum technologies in hBN.

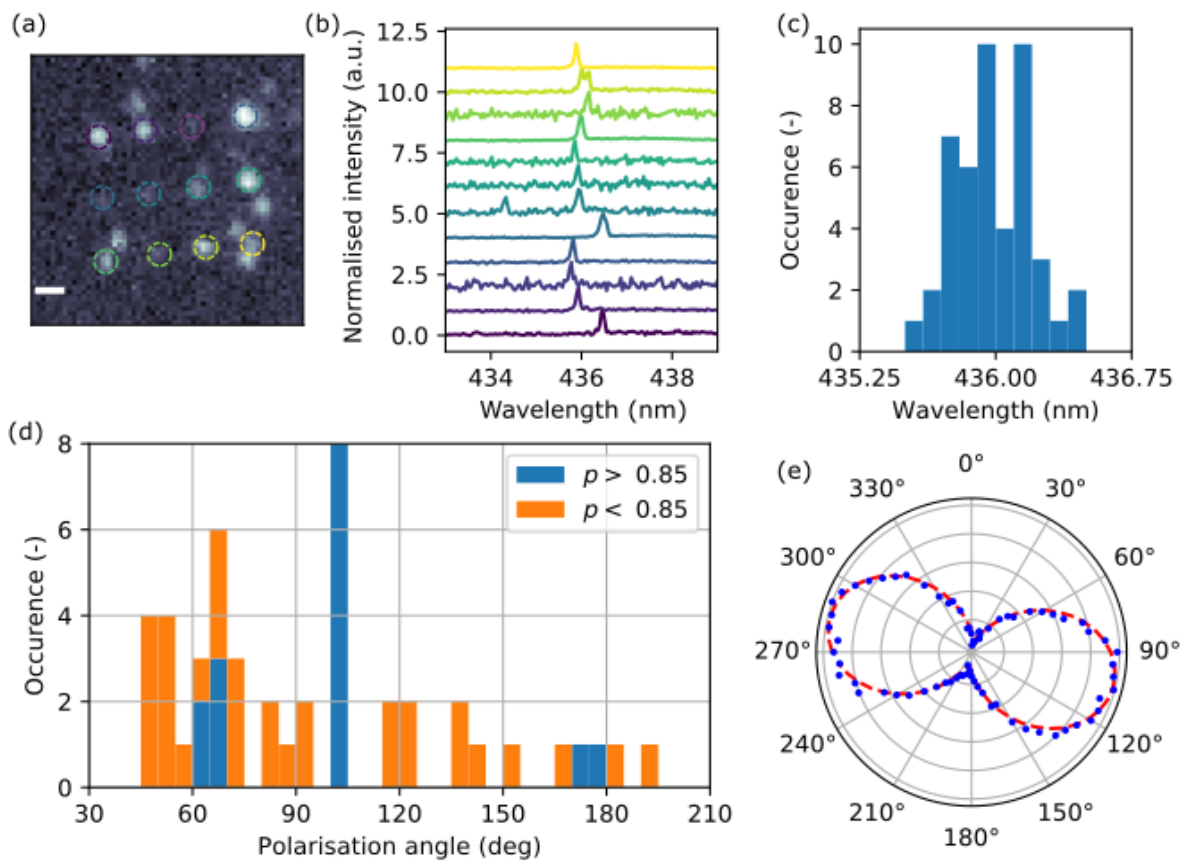

*Fig. 1. Emission characterization. a) Confocal PL scan of the 12 sites in Matrix A (MA), one of three emitter arrays investigated in this work. Scale bar 1 μm. b) Normalized PL spectra measured at each site circled in (a). Each plot is vertically offset for clarity. c) Histogram of ZPL wavelengths extracted from PL spectra across 46 emitter sites at three separate locations on the hBN flake. The mean wavelength is 435.98 nm ± 0.21 nm. d) Histogram of emission polarization angles measured from 46 sites, segregated by a polarisability threshold of 0.85. e) A typical polarization measurement with data in blue and a sinusoidal fit in dashed red.*

**Results**

Spectral instability such as spectral diffusion and blinking of quantum emitters in hBN are known issues hindering practical application of such emitters[27,28]. The recently-discovered blue emitter in hBN, generated by electron beam irradiation, was shown to have a stable emission intensity at around 436 nm. To investigate the emission properties of these emitters in detail, a flake of hBN hosting blue emitters created using electron beam irradiation was mounted in a cryostat and cooled to 5 K. The fabrication process of these emitters[26] is explained in the methods section. Three arrays of 3-by-4 sites were patterned on a single flake for this study. To reveal the locations of the emitters, a confocal photoluminescence (PL) scan was performed over a 10 μm × 10 μm area containing each array. The arrays are here termed Matrix A (MA), Matrix B (MB), and Matrix C (MC). These confocal scans were performed using a 405 nm laser operating at 1 mW, with collection filtered using a 436 nm ± 3 nm band pass filter. Figure 1(a) shows the confocal PL scan over MA, and confocal scans of the other arrays are presented in the Supplementary Information (SI). Measurements of the high resolution PL emission spectrum from each site in MA are shown collectively in Fig. 1(b). Each spectrum has been normalized and offset vertically for clarity. The twin emission maxima evident in the spectrum plotted second from the top is most likely from the presence of multiple emitters at this site. Time series of PL spectra revealed no discernable temporal variation in intensity (blinking) or peak wavelength (spectral diffusion), as is seen in Fig. S1(d). PL spectra were also measured for sites located in MB and MC, and the distribution of peak emission wavelengths is shown in Fig. 1(c). Across the 46 sites measured, the average zero phonon line (ZPL) wavelength is 435.98 nm ± 0.21 nm. The small variability in emission wavelength reflects a high degree of consistency in the spectral properties of this class of emitter, which we attribute to preferential fabrication of a specific defect structure by the electron beam, and a high purity and crystallinity of the hBN sample grown by the high pressure high temperature (HPHT) method. We note that this narrow distribution of emission wavelength is uncommon in other classes of emitters in hBN[9,29].

Previous work has speculated on the nature of the SPE formation under electron beam irradiation and the resulting defect structure which may feature substitutional, vacancy, or interstitial atoms[24-26]. Structural insight may be gained indirectly through measurements of the emission dipole direction and polarisability. With this in mind, polarization measurements were performed at each of the available sites with the aim of observing a potential correlation between the most common polarization directions and the orientation of the underlying hBN crystal lattice which has three-fold rotational symmetry.

To study the emission polarization, emitters were illuminated with a circularly polarized laser at 405 nm. The polar plot in Fig. 1(e) shows a polarization measurement typical of the sites in MA. Further polar plots in Fig. S2(b) show the results of polarization measurements performed at each site in MA. The data are normalized and fit with a cosine function:

$$p = \alpha \cos^2(\theta - \theta_0) + \beta$$

The polarisability is defined as the difference between the maximum and minimum values of the fitted function. Polarization measurements were performed on all 46 sites located on the same flake, and the histogram in Fig. 1(d) shows the distribution of polarization angles observed. The bin size is 5 degrees, and the data are segregated according to an arbitrary polarisability threshold of p = 0.85. Although there is clustering for angles corresponding to a high degree of polarisability, there is no clear correlation between the lattice symmetry angle and the clustering of polarization directions, consistent with the result from Fournier et al[25]. The polarisability threshold was used in an attempt to filter out the low polarizability caused by the presence of multiple emitters in one site. There appears to be no strong correlation between the polarisability and the single photon purity as determined by the second order correlation function (see SI, Fig. S2). That is, sites with relatively low polarisability have also been found to be single photon emitters. These results may indicate that the studied defects are either substitutional or interstitial defects without a clear symmetry axis, unlike the proposed carbon trimers for the visible quantum emitters[30,31].

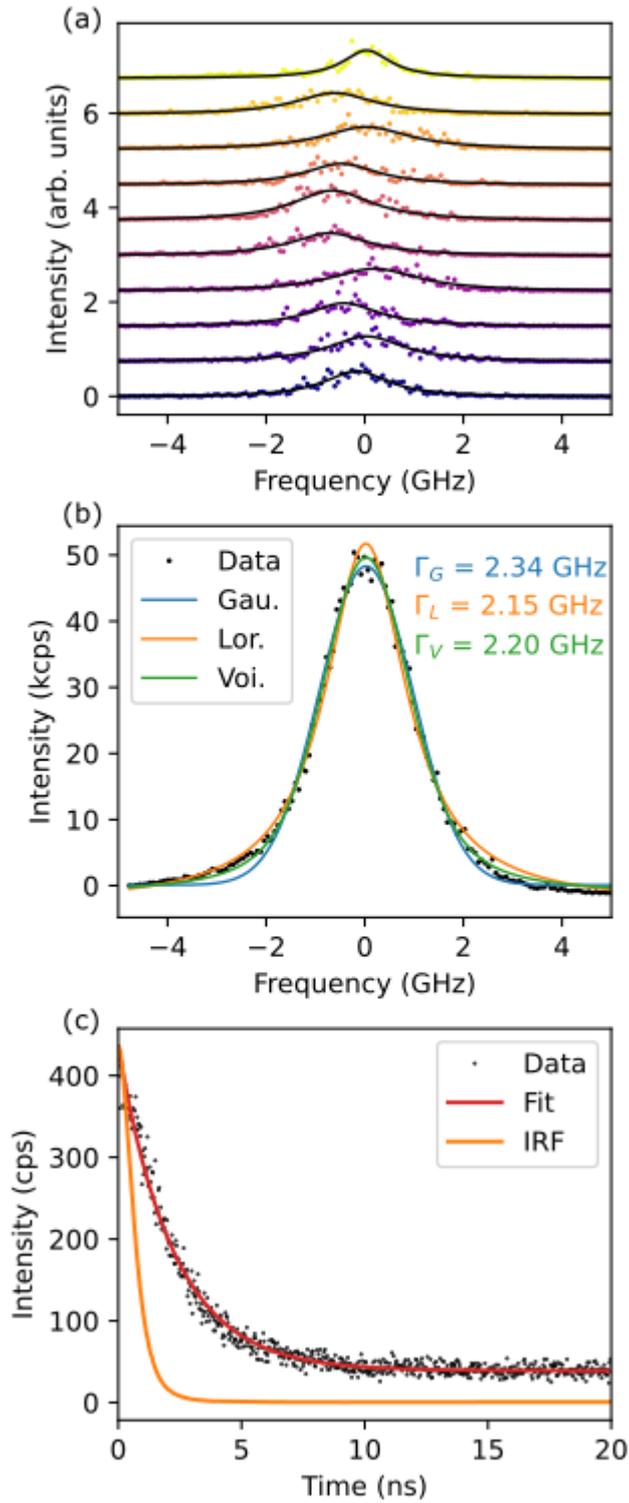

Fig. 2. Photoluminescence excitation. a) Ten consecutive PLE scans across the ZPL of an emitter reveal some spectral wandering. Intensity data from each scan is fit with a Lorentzian function and offset vertically for clarity. b) Average fluorescence intensity from 50 consecutive scans. These data are fit with a Gaussian function, a Lorentzian function, and a Voigt function, from which estimates of the linewidth $\Gamma$ (FWHM) are extracted. c) Lifetime measurement obtained using a pulsed laser. The exponential fit yields the excited state lifetime $T_1$ = 2.27 ns ± 0.02 ns. The orange curve is the instrument response function (IRF).

To explore spectral properties of these emitters beyond the resolution limit of the PL spectra in Fig. 1(b), resonant photoluminescence excitation (PLE) measurements were performed. A Ti:Sap laser was used to scan across the ZPL of an emitter, with the resulting fluorescence collected from the phonon sideband (PSB). In Fig. 2(a), the data sets show the fluorescence intensity from ten consecutive scans across the ZPL, with each scan offset vertically for clarity. The data from each scan are fit with a Lorentzian function, and for 50 consecutive scans the average linewidth is 1.73 GHz, taken by averaging the full width at half maximum (FWHM) of each fitted curve. The average ZPL was approximately 435.745 nm, or 688017 GHz. The consecutive fits also indicate a slight degree of spectral diffusion, as shown by the frequency variation at which the peak intensity occurs. The set of ZPLs is approximately normally distributed, and fitting a Gaussian function to this distribution yields an FWHM of 0.91 GHz, representing an estimate of the extent of inhomogeneous broadening.

Because the magnitude of spectral diffusion is less than the average single-scan linewidth, the aggregate lineshape after many scans remains predominantly Lorentzian. This is shown in Fig. 2(b), where the average intensity over a given frequency interval for all 50 consecutive scans is plotted, together with both a Lorentzian fit and a Gaussian fit that respectively model homogeneous and inhomogeneous linewidth broadening. A Voigt function (i.e., a convolution of a Gaussian and a Lorentzian function) fit is also shown in Fig. 2(b). Each function yields a similar linewidth, shown as inset values. However an analysis of the residuals from each fit shows that the Voigt function gives the least error, and hence we conclude both homogeneous and inhomogeneous mechanisms are influencing the total linewidth. The Voigt fit allows the linewidths of the constituent functions to be extracted, and for these data the contributions to the convolution are $\Gamma_G$ = 1.50 GHz for the Gaussian component, and $\Gamma_L$ = 1.06 GHz for the Lorentzian component, which are close to the values obtained from the analysis of the 50 consecutive scans.

The emission linewidth in the absence of broadening processes is determined by the lifetime of the excited state, $T_1$, through the uncertainty principle. Emitter lifetime was measured using a pulsed 405 nm laser operating at 100 μW with a pulse frequency of 20 MHz. From an exponential fit to the data, shown in Fig. 2(c), $T_1$ was determined to be 2.27 ns ± 0.02 ns, which is comparable to the lifetime of other emitter classes in hBN[24–26,29]. Background emission was measured at a location on the hBN flake away from the emitter and used for a background corrected fit. A measurement of the instrument response function (IRF) is also shown in Fig. 2(c). The IRF data have been normalized with respect to the maximum intensity observed over the emitter for ease of comparison. The lifetime measurement yields an estimate of 70 MHz for the natural linewidth of this emitter. This value is much less than the linewidths obtained from the PLE scans above, confirming the presence of homogeneous and/or inhomogeneous broadening mechanisms.

The lineshape observed from PLE scans became more regular with increasing excitation power. Several scans across the ZPL of an emitter were performed for powers ranging from 0.04 μW to 2.4 μW. Figure 3(a) shows the PLE data collected at each excitation power, with data sets for increasing power offset vertically. For each data set, the measured emission intensity was fit with a Voigt function. The same analysis undergone above indicates that this function is a better model for the intensity distribution than either the Gaussian or Lorentzian functions. The Voigt fit allows a quantification of the maximum intensity and linewidth of the ZPL, and in Fig. 3(b) the maximum intensity is plotted for each excitation power. These maxima display saturation behavior for increasing power and can be adequately described by the function,

$$I = \frac{I_\infty P}{P + P_0}$$

where $I$ is the emission intensity, $I_\infty$ is the maximum obtainable intensity, $P$ is the excitation power, and $P_0$ is the saturation power. For this emitter, a saturation power of 3.14 µW is extracted from the fitted function. Increasing excitation power up to saturation does not appear to have a strong influence on the linewidth. Figure 3(c) shows the full width at half maximum (FWHM) of each Voigt fit plotted in Fig. 3(a), with the average linewidth found to be 2.13 GHz.

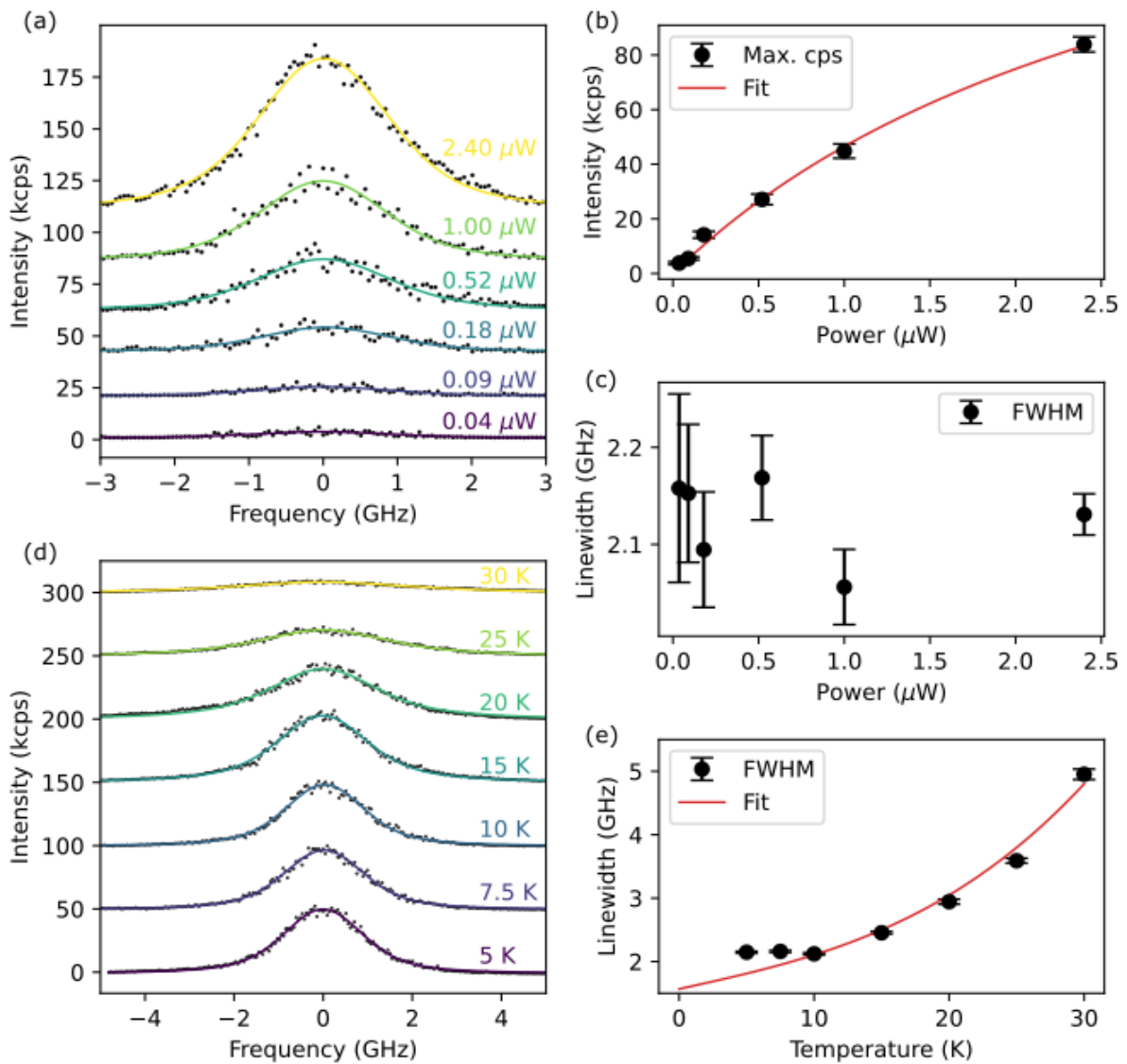

Fig. 3. Linewidth broadening. a) PLE collected over a range of excitation powers. Each dataset is fit with a Voigt function and plotted with a vertical offset for clarity. b) The maximum value of each Voigt function in (a) is plotted against the

*excitation power, and fitted with a power-intensity function to extract the saturation power 3.14 µW. c) The full width at half maximum of each fitted function in (a) is plotted against the excitation power. d) PLE collected at a range of temperatures, using an excitation power of 1 µW. Each dataset is fit with a Voigt function and plotted with a vertical offset for clarity. e) The FWHM of each fitted function in (d) is plotted against the temperature and fit using a cubic function.*

To investigate the influence of temperature on the emission spectrum of the emitter, PLE scans were performed using 1 µW laser power, over 5 minutes of data collection, for a range of temperatures between 5 K and 30 K. The lineshape was observed to widen in proportion to temperature, such that above 30 K the width had exceeded the maximum scanning interval of the laser, hence no data were collected past 30 K. The magnitude of inhomogeneous broadening was estimated by extracting the Gaussian component of the Voigt fit to the data at 5 K. This value of $\Gamma_G = 1.50$ GHz was then fixed for subsequent Voigt fits. The data sets and their Voigt fits are shown in Fig. 3(d), offset vertically with increasing temperature. Figure 3(e) shows the increasing FWHM together with a fitting function based on the model[32–34]

$$\Gamma = \alpha T^3 + \beta T + \Gamma_0$$

where $\Gamma$ is the temperature-broadened linewidth, $\Gamma_0$ is the temperature-independent linewidth, and $T$ is the temperature. The fit assumes that homogeneous broadening vanishes at 0 K, and hence only the natural linewidth of 70 MHz and inhomogeneous broadening of 1.50 GHz account for a fixed value of $\Gamma_0 = 1.57$ GHz. We attribute the discrepancy between the fit and the linewidth observed at 5 K and 7.5 K to additional broadening due to local heating[35,36].

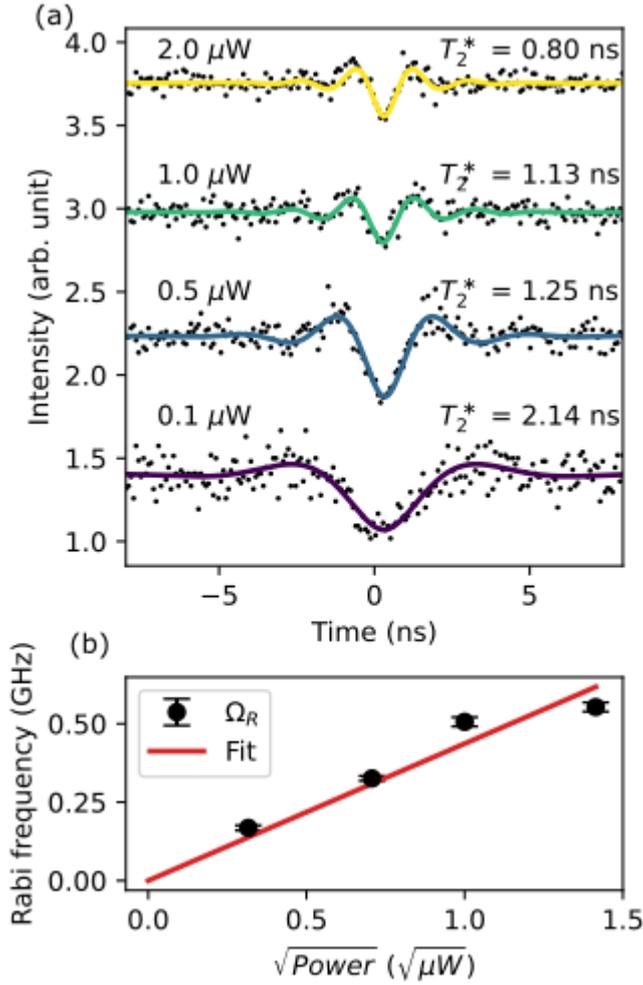

Fig. 4. Coherence properties. a) Second order correlation measurement collected over a range of excitation powers, and the fitted function demonstrating the presence of Rabi oscillations. Each plot is vertically offset for clarity. Values of $T_2^*$ are extracted from the fit. b) The Rabi frequency extracted from the fit in (a) is plotted against the square root of the excitation power, and a linear function is used to fit the data.

When exciting at low power, the transition from the ground state to excited state occurs through spontaneous absorption of an incident photon from the excitation source. However when the excitation laser is operating at high power, the emitter is driven to a coherent superposition of ground and excited states. Further, continuous high power excitation modulates the probability amplitudes in a cyclic way, known as Rabi oscillation. The modulation of the excited state probability is reflected in the bunching statistics of consecutive photons released by the emitter. Hence, Rabi oscillations are observable in measurements of $g^{(2)}$, the second order correlation function. The Ti:Sap laser was used to resonantly excite an emitter at powers in the range of 0.1 μW to 2 μW. The collected emission was then directed to a fiber based Hanbury-Brown and Twiss (HBT) setup. Figure 4(a) shows the second order correlation at each power, using 32 ps bins and a total duration of 45 minutes. Each data set is fit with the exponential oscillation function[37]

$$g^{(2)}(\tau) = 1 - \exp(-\eta|\tau - \tau_0|)\left(\frac{\eta}{\Omega}\sin(\Omega|\tau - \tau_0|) + \cos(\Omega|\tau - \tau_0|)\right)$$

where $\eta = 3/(4T_1) + 1/(2T_2^*)$, $T_1$ is the excited state lifetime, $T_2^*$ is the pure dephasing time, and $\Omega_R/2\pi$ is the Rabi frequency. The value of $T_1$ was fixed based on the lifetime measurement detailed above, while $T_2^*$ and $\Omega_R$ were fitting parameters. The pure dephasing time decreases with increasing power, according to the fitted values shown inset above each data set in Fig. 4(a). The Rabi frequency should be directly proportional to the magnitude of the applied electric field, so that there is a square root relation between incident laser intensity and observed oscillation frequency. Figure 4(b) shows that the Rabi frequency is well described by a linear fit to the square root of the excitation laser power, as expected. These measurements were performed at two other emitter sites, one of which similarly yielded Rabi oscillations (see Fig. S4).

For the case of 1 uW resonant excitation, the pure dephasing time obtained from the fit ($T_2^*$ = 1.13 ns) gives a coherence time $T_2$ of 0.90 ns[38]. This describes the coherence time for two consecutive photons and gives a $2T_1/T_2$ ratio of ~ 5. This is much greater than the ideal ratio of 1, suggesting that two photon interference performed on this current system would yield far from perfect indistinguishability[39]. The coherence time is likely reduced further under the influence of phonon interactions and spectral diffusion that cause linewidth broadening[40], which would be relevant for the interference of photons with significant temporal separation. Nevertheless, Reimer et al[36] note that Fourier-limited photons are not essential for observing indistinguishability, so long as the instrument response time is much shorter than the coherence time. This leaves the door open for future two photon interference experiments with these blue emitters, which would be served in the meantime by complementary experiments aiming to quantify and control the rate of spectral diffusion[41,42].

**Conclusion**

This work represents the first in depth study of a new hBN defect at cryogenic temperature. The blue emitters can be reliably fabricated to the extent that we were able to gather statistics on the spectral properties of several dozen emitters. In particular, we were able to achieve Rabi oscillations in autocorrelation functions measured from two different emitters. Power and temperature broadening measurements indicate that the linewidth is predominantly influenced by phonon-assisted recombination, although the relative contribution from spectral diffusion is less clear. Furthermore, lack of knowledge of the energy level structure for this class of emitter invokes interest in the role of possible metastable states, and future work should consider the power dependence of the second order correlation function to explore this feature. The observation of Rabi oscillations demonstrates that the defect associated with this blue emitter is capable of supporting a coherent superposition state, which further enhances the appeal of hBN as a platform for quantum technologies.

**Methods**

Sample preparation

Bulk hBN grown using the High Pressure High Temperature (HPHT) method was used to exfoliate a thin hBN flake onto a $SiO_2$/Si substrate, followed by an annealing treatment in $N_2$ at 1000 °C. An isolated flake on the sample was then subject to electron beam irradiation using a 10kV beam, 1nA current and 2s irradiation time per spot, resulting in the controlled formation of fluorescent defects at chosen sites on the flake. A total of three arrays of 3 ✕ 4 defect sites were patterned onto the flake. The sample was then cooled to 5 K using a closed-loop cryostat.

Photoluminescence experiments

PL spectra were measured using an Andor Spectrometer with 1800 lines/mm grating, integrating for 5 s under 1 mW of 405 nm laser excitation. To measure the emission polarization from each site the sample was illuminated with a circularly polarised 1 mW 405 nm laser. The fluorescence from each emitter site was directed through a rotatable ½ λ waveplate (HWP) and onto a polarizing beam splitter. PL counts were then measured as the HWP was rotated through 180°. The lifetime measurement was performed using a pulsed 405 nm laser operating at 100 μW with a pulse frequency of 20 MHz. Emission was also measured at a location on the hBN flake away from the emitter in order to perform background correction. For all PLE experiments, resonant excitation was performed with a Ti:sapphire scanning laser (M Squared) and photons were collected from the phonon side band (PSB). Data were collected over 25 ms integration time. Scanning was performed at a rate of 1 GHz/s unless otherwise specified. Time series of PLE spectra were taken over 280 s at 1 μW. Second order correlation measurements were performed using APDs and a time tagging device (Swabian).

**Acknowledgements**

The authors acknowledge financial support from the Australian Research Council (CE200100010, DP190101058) and the Office of Naval Research Global (N62909-22-1-2028) for financial support. The authors thank the UTS node of Optofab ANFF for the assistance with nanofabrication. This research is supported by an Australian Government Research Training Program Scholarship. Kenji Watanabe and Takashi Taniguchi acknowledge support from the Elemental Strategy Initiative conducted by the MEXT, Japan, Grant No. JPMXP0112101001, JSPS KAKENHI Grant No. JP20H00354 and the CREST (JPMJCR15F3), JST.

**Disclosures**

The authors declare no conflicts of interest.

**References**

1. Levine EV, Turner MJ, Kehayias P, Hart CA, Langellier N, Trubko R, Glenn DR, Fu RR, Walsworth RL. Principles and techniques of the quantum diamond microscope. *Nanophotonics*. 2019;8(11):1945-1973.
2. Wehner S, Elkouss D, Hanson R. Quantum internet: A vision for the road ahead. *Science*. 2018;362(6412):eaam9288.
3. Pezzagna S, Meijer J. Quantum computer based on color centers in diamond. *Appl Phys Rev*. 2021;8(1):011308.
4. Elshaari AW, Pernice W, Srinivasan K, Benson O, Zwiller V. Hybrid integrated quantum photonic circuits. *Nat Photonics*. 2020;14(5):285-298.
5. Awschalom DD, Hanson R, Wrachtrup J, Zhou BB. Quantum technologies with optically interfaced solid-state spins. *Nat Photonics*. 2018;12(9):516-527.
6. Humphreys PC, Kalb N, Morits JPJ, Schouten RN, Vermeulen RFL, Twitchen DJ, Markham M, Hanson R. Deterministic delivery of remote entanglement on a quantum network. *Nature*. 2018;558(7709):268-273.
7. Sipahigil A, Evans RE, Sukachev DD, Burek MJ, Borregaard J, Bhaskar MK, Nguyen

CT, Pacheco JL, Atikian HA, Meuwly C, Camacho RM, Jelezko F, Bielejec E, Park H, Lončar M, Lukin MD. An integrated diamond nanophotonics platform for quantum-optical networks. *Science*. 2016;354(6314):847-850.
8. Exarhos AL, Hopper DA, Patel RN, Doherty MW, Bassett LC. Magnetic-field-dependent quantum emission in hexagonal boron nitride at room temperature. *Nat Commun*. 2019;10(1):222.
9. Gottscholl A, Kianinia M, Soltamov V, Orlinskii S, Mamin G, Bradac C, Kasper C, Krambrock K, Sperlich A, Toth M, Aharonovich I, Dyakonov V. Initialization and read-out of intrinsic spin defects in a van der Waals crystal at room temperature. *Nat Mater*. 2020;19(5):540-545.
10. Gottscholl A, Diez M, Soltamov V, Kasper C, Sperlich A, Kianinia M, Bradac C, Aharonovich I, Dyakonov V. Room temperature coherent control of spin defects in hexagonal boron nitride. *Sci Adv*. 2021;7(14).
11. Guo NJ, Liu W, Li ZP, Yang YZ, Yu S, Meng Y, Wang ZA, Zeng XD, Yan FF, Li Q, Wang JF, Xu JS, Wang YT, Tang JS, Li CF, Guo GC. Generation of Spin Defects by Ion Implantation in Hexagonal Boron Nitride. *ACS Omega*. 2022;7(2):1733-1739.
12. Stern HL, Gu Q, Jarman J, Eizagirre Barker S, Mendelson N, Chugh D, Schott S, Tan HH, Sirringhaus H, Aharonovich I, Atatüre M. Room-temperature optically detected magnetic resonance of single defects in hexagonal boron nitride. *Nat Commun*. 2022;13(1):618.
13. Gao X, Pandey S, Kianinia M, Ahn J, Ju P, Aharonovich I, Shivaram N, Li T. Femtosecond Laser Writing of Spin Defects in Hexagonal Boron Nitride. *ACS Photonics*. 2021;8(4):994-1000.
14. Bourrellier R, Meuret S, Tararan A, Stéphan O, Kociak M, Tizei LHG, Zobelli A. Bright UV Single Photon Emission at Point Defects in h-BN. *Nano Lett*. 2016;16(7):4317-4321.
15. Kianinia M, Regan B, Tawfik SA, Tran TT, Ford MJ, Aharonovich I, Toth M. Robust Solid-State Quantum System Operating at 800 K. *ACS Photonics*. 2017;4(4):768-773.
16. Mendelson N, Chugh D, Reimers JR, Cheng TS, Gottscholl A, Long H, Mellor CJ, Zettl A, Dyakonov V, Beton PH, Novikov SV, Jagadish C, Tan HH, Ford MJ, Toth M, Bradac C, Aharonovich I. Identifying carbon as the source of visible single-photon emission from hexagonal boron nitride. *Nat Mater*. 2021;20(3):321-328.
17. Malein RNE, Khatri P, Ramsay AJ, Luxmoore IJ. Stimulated Emission Depletion Spectroscopy of Color Centers in Hexagonal Boron Nitride. *ACS Photonics*. 2021;8(4):1007-1012.
18. Nikolay N, Mendelson N, Özelci E, Sontheimer B, Böhm F, Kewes G, Toth M, Aharonovich I, Benson O. Direct measurement of quantum efficiency of single-photon emitters in hexagonal boron nitride. *Optica*. 2019;6(8):1084-1088.
19. Li C, Fröch JE, Nonahal M, Tran TN, Toth M, Kim S, Aharonovich I. Integration of hBN Quantum Emitters in Monolithically Fabricated Waveguides. *ACS Photonics*. 2021;8(10):2966-2972.
20. Groll D, Hahn T, Machnikowski P, Wigger D, Kuhn T. Controlling photoluminescence spectra of hBN color centers by selective phonon-assisted excitation: a theoretical proposal. *Mater Quantum Technol*. 2021;1(1):015004.
21. Konthasinghe K, Chakraborty C, Mathur N, Qiu L, Mukherjee A, Fuchs GD, Vamivakas AN. Rabi oscillations and resonance fluorescence from a single hexagonal boron nitride quantum emitter. *Optica*. 2019;6(5):542-548.
22. Atatüre M, Englund D, Vamivakas N, Lee SY, Wrachtrup J. Material platforms for spin-based photonic quantum technologies. *Nat Rev Mater*. 2018;3(5):38-51.
23. Sontheimer B, Braun M, Nikolay N, Sadzak N, Aharonovich I, Benson O. Photodynamics of quantum emitters in hexagonal boron nitride revealed by low-temperature spectroscopy. *Phys Rev B*. 2017;96(12):121202.
24. Shevitski B, Gilbert SM, Chen CT, Kastl C, Barnard ES, Wong E, Ogletree DF, Watanabe K, Taniguchi T, Zettl A, Aloni S. Blue-light-emitting color centers in high-quality hexagonal boron nitride. *Phys Rev B*. 2019;100(15):155419.

25. Fournier C, Plaud A, Roux S, Pierret A, Rosticher M, Watanabe K, Taniguchi T, Buil S, Quélin X, Barjon J, Hermier JP, Delteil A. Position-controlled quantum emitters with reproducible emission wavelength in hexagonal boron nitride. *Nat Commun*. 2021;12(1):3779.
26. Gale A, Li C, Chen Y, Watanabe K, Taniguchi T, Toth M. Deterministic fabrication of blue quantum emitters in hexagonal boron nitride. *ArXiv Prepr*. 2021; arXiv:2111.13441.
27. Tran TT, Kianinia M, Nguyen M, Kim S, Xu ZQ, Kubanek A, Toth M, Aharonovich I. Resonant Excitation of Quantum Emitters in Hexagonal Boron Nitride. *ACS Photonics*. 2018;5(2):295-300.
28. Kianinia M, Bradac C, Sontheimer B, Wang F, Tran TT, Nguyen M, Kim S, Xu ZQ, Jin D, Schell AW, Lobo CJ, Aharonovich I, Toth M. All-optical control and super-resolution imaging of quantum emitters in layered materials. *Nat Commun*. 2018;9(1):874.
29. Tran TT, Bray K, Ford MJ, Toth M, Aharonovich I. Quantum emission from hexagonal boron nitride monolayers. *Nat Nanotechnol*. 2016;11(1):37-41.
30. Golami O, Sharman K, Ghobadi R, Wein SC, Zadeh-Haghighi H, Gomes da Rocha C, Salahub DR, Simon C. $Ab initio$ and group theoretical study of properties of a carbon trimer defect in hexagonal boron nitride. *Phys Rev B*. 2022;105(18):184101.
31. Li K, Smart TJ, Ping Y. Carbon trimer as a 2 eV single-photon emitter candidate in hexagonal boron nitride: A first-principles study. *Phys Rev Mater*. 2022;6(4):L042201.
32. White S, Stewart C, Solntsev AS, Li C, Toth M, Kianinia M, Aharonovich I. Phonon dephasing and spectral diffusion of quantum emitters in hexagonal boron nitride. *Optica*. 2021;8(9):1153.
33. Hizhnyakov V, Kaasik H, Sildos I. Zero-Phonon Lines: The Effect of a Strong Softening of Elastic Springs in the Excited State. *Phys Status Solidi B*. 2002;234(2):644-653.
34. Jahnke KD, Sipahigil A, Binder JM, Doherty MW, Metsch M, Rogers LJ, Manson NB, Lukin MD, Jelezko F. Electron–phonon processes of the silicon-vacancy centre in diamond. *New J Phys*. 2015;17(4):043011.
35. Li X, Shepard GD, Cupo A, Camporeale N, Shayan K, Luo Y, Meunier V, Strauf S. Nonmagnetic Quantum Emitters in Boron Nitride with Ultranarrow and Sideband-Free Emission Spectra. *ACS Nano*. 2017;11(7):6652-6660.
36. Reimer ME, Bulgarini G, Fognini A, Heeres RW, Witek BJ, Versteegh MAM, Rubino A, Braun T, Kamp M, Höfling S, Dalacu D, Lapointe J, Poole PJ, Zwiller V. Overcoming power broadening of the quantum dot emission in a pure wurtzite nanowire. *Phys Rev B*. 2016;93(19):195316.
37. Batalov A, Zierl C, Gaebel T, Neumann P, Chan IY, Balasubramanian G, Hemmer PR, Jelezko F, Wrachtrup J. Temporal Coherence of Photons Emitted by Single Nitrogen-Vacancy Defect Centers in Diamond Using Optical Rabi-Oscillations. *Phys Rev Lett*. 2008;100(7):077401.
38. Spokoyny B, Utzat H, Moon H, Grosso G, Englund D, Bawendi MG. Effect of Spectral Diffusion on the Coherence Properties of a Single Quantum Emitter in Hexagonal Boron Nitride. *J Phys Chem Lett*. 2020;11(4):1330-1335.
39. Huber T, Predojević A, Föger D, Solomon G, Weihs G. Optimal excitation conditions for indistinguishable photons from quantum dots. *New J Phys*. 2015;17(12):123025.
40. Nawrath C, Olbrich F, Paul M, Portalupi SL, Jetter M, Michler P. Coherence and indistinguishability of highly pure single photons from non-resonantly and resonantly excited telecom C-band quantum dots. *Appl Phys Lett*. 2019;115(2):023103.
41. Buckley S, Rivoire K, Vučković J. Engineered quantum dot single-photon sources. *Rep Prog Phys*. 2012;75(12):126503.
42. Wolters J, Sadzak N, Schell AW, Schröder T, Benson O. Measurement of the Ultrafast Spectral Diffusion of the Optical Transition of Nitrogen Vacancy Centers in Nano-Size Diamond Using Correlation Interferometry. *Phys Rev Lett*. 2013;110(2):027401.

# Supplementary Information
# Coherence properties of electron beam activated emitters in hexagonal boron nitride under resonant excitation

Jake Horder[1], Simon J. U. White[1], Angus Gale[1], Chi Li[1], Kenji Watanabe[2], Takashi Taniguchi[3], Mehran Kianinia[1,4], Igor Aharonovich[1,4] and Milos Toth[1,4]

1. School of Mathematical and Physical Sciences, University of Technology Sydney, Ultimo, New South Wales 2007, Australia

2. Research Center for Functional Materials, National Institute for Materials Science, Tsukuba 305-0044, Japan

3. International Center for Materials Nanoarchitectonics, National Institute for Materials Science, Tsukuba 305-0044, Japan

4. ARC Centre of Excellence for Transformative Meta-Optical Systems, University of Technology Sydney, Ultimo, New South Wales 2007, Australia

milos.toth@uts.edu.au; igor.aharonovich@uts.edu.au

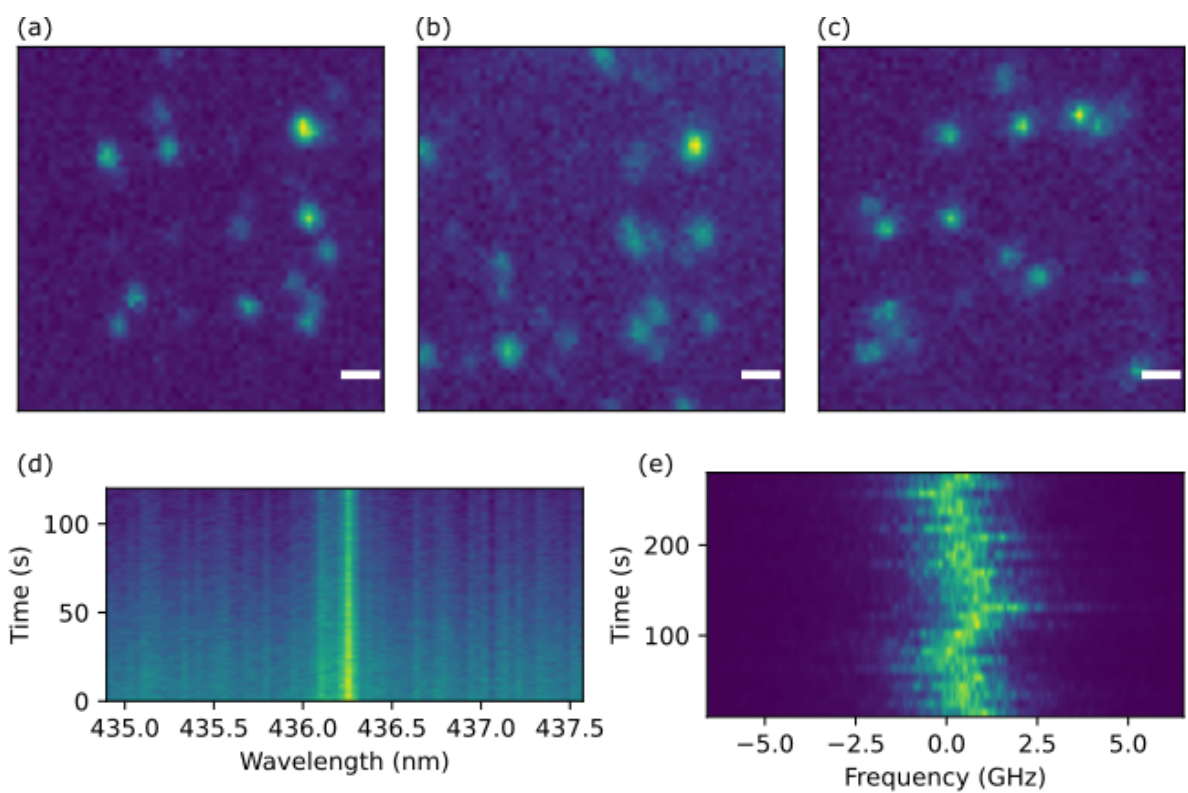

Fig. S1. Spatial and spectral characterisation. a) Confocal scans of region MA. Scale bar 1 µm. b) Confocal scans of region MB. Scale bar 1 µm. c) Confocal scans of region MC. Scale bar 1 µm. d) PL time series using 1800 lines/mm grating over 120 s, 405 nm laser excitation at 1 mW. e) PLE time series scanning 1 GHz/s over 240 s.

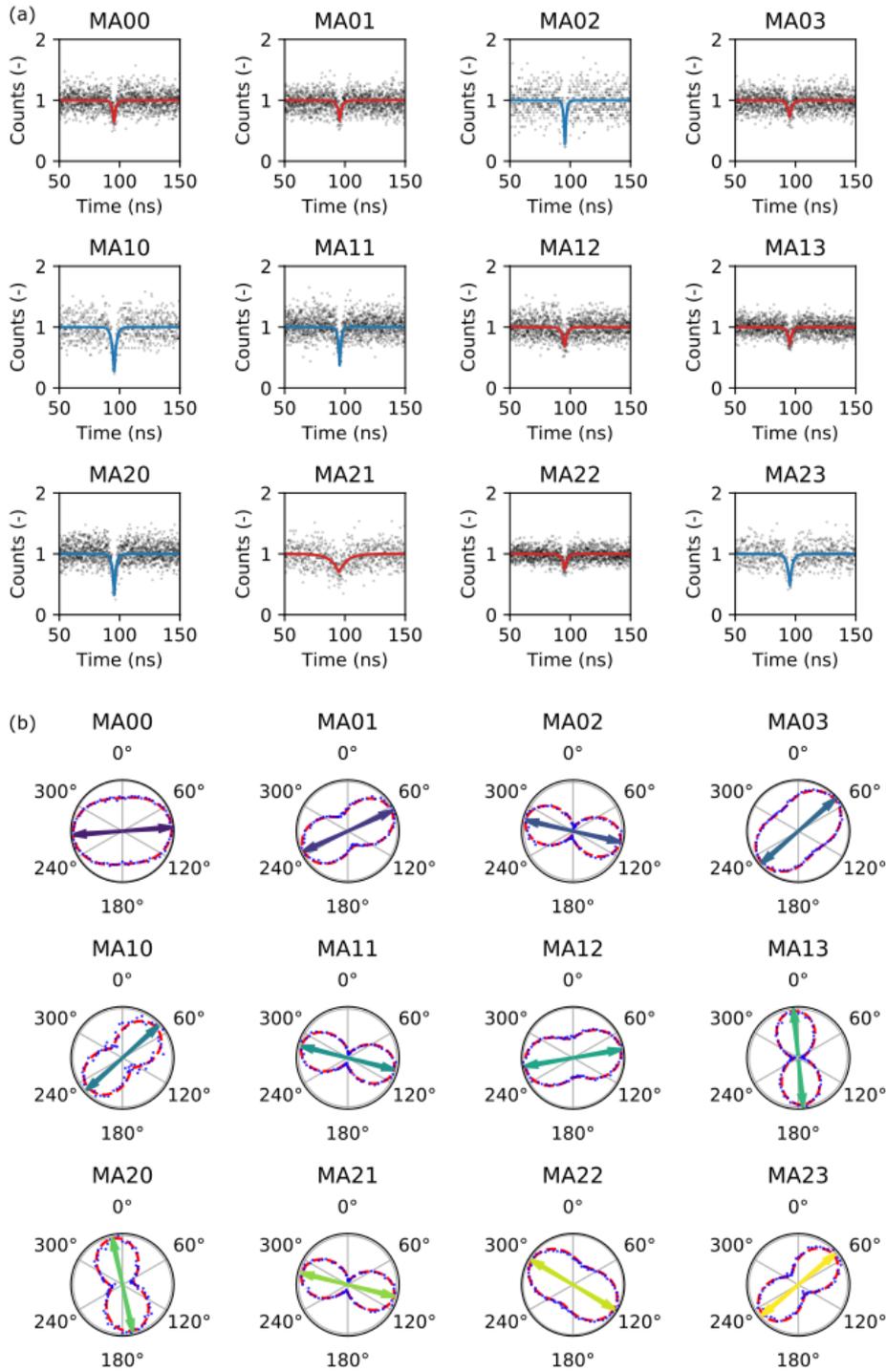

Fig. S2. Emitter purity and polarisation. a) Second order correlation measurements at each emitter site in MA, performed at room temperature. Fitted plots in blue have $g^{(2)} < 0.5$, while those in red have $g^{(2)} > 0.5$. These data sets have not been background corrected. b) Polar plots of the emission polarisation collected over each site in MA. Each plot is fitted with a function of the form $\alpha \cos^2(\theta - \theta_0) + \beta$, and the primary polarisation axis is emphasised with a double headed arrow.

**Voigt function**

Given a random variable $X$ sampled from a Lorentzian probability distribution $L$, and a second independent random variable $Y$ sampled from a Gaussian probability distribution $G$, then the random variable defined as $Z = X + Y$ will have a probability distribution given by $V = L * G$, where $V$ is the Voigt function and * is the convolution operation. In practice, the Voigt function is computed using the expression

$$V = \frac{\Re\left[W\left(\frac{x+if_L/2}{(f_G/2)/2\sqrt{2\ln(2)}}\right)\right]}{(f_G/2)/2\sqrt{2\ln(2)}}$$

where $f_L$ is the FWHM of the component distribution $L$, $f_G$ is the FWHM of the component distribution $G$, $R[W]$ is the real part of the Faddeeva function. The FWHM of the Voigt function can be approximated by

$$f_V \approx 0.5346 f_L + \sqrt{0.2166 f_L^2 + f_G^2}$$

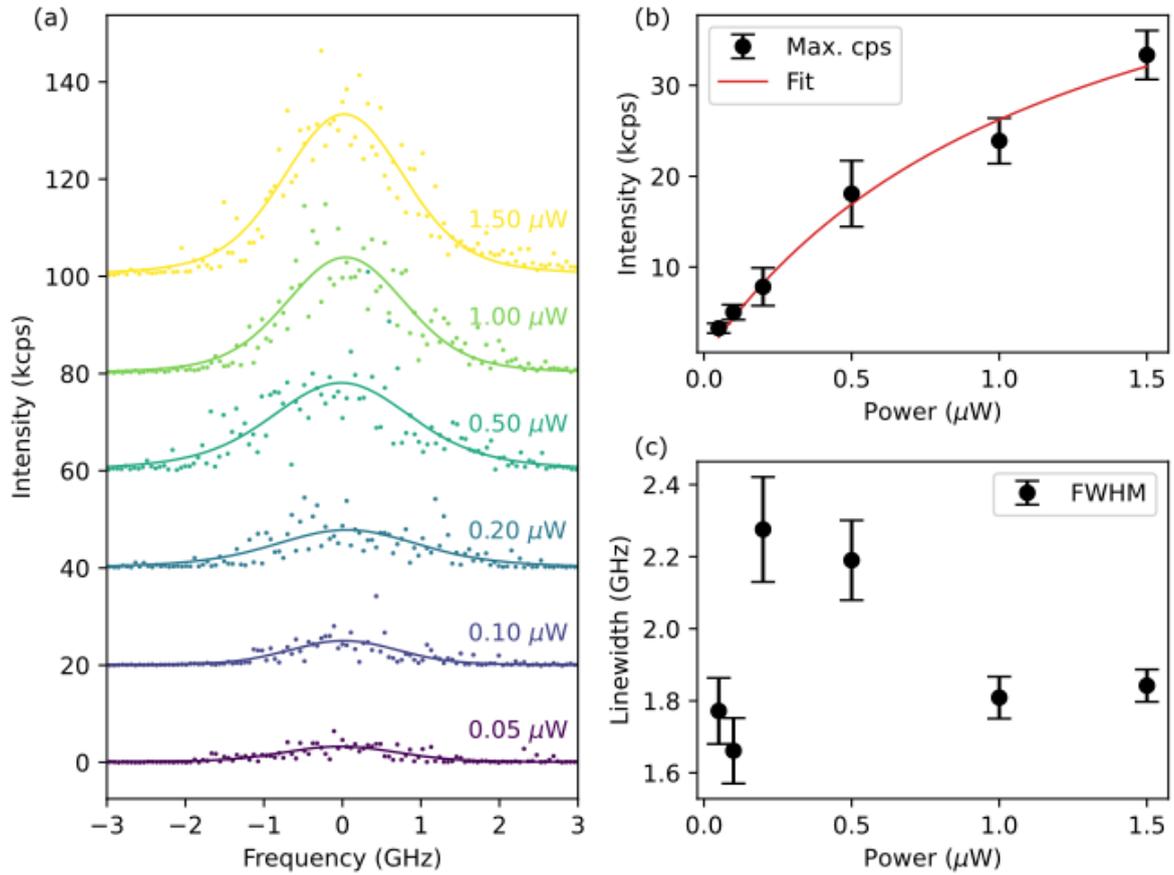

Fig. S3. Power saturation for site in MA. a) PLE scans for a range of excitation powers, with each data set offset vertically for clarity. Lines are Voigt function fits to the data. b) Maximum value from fitted functions in (a) plotted against excitation power. The fitted curve gives a saturation power of 1.21 µW. c) Linewidth of each fitted function in (a), with an average of 1.92 GHz.

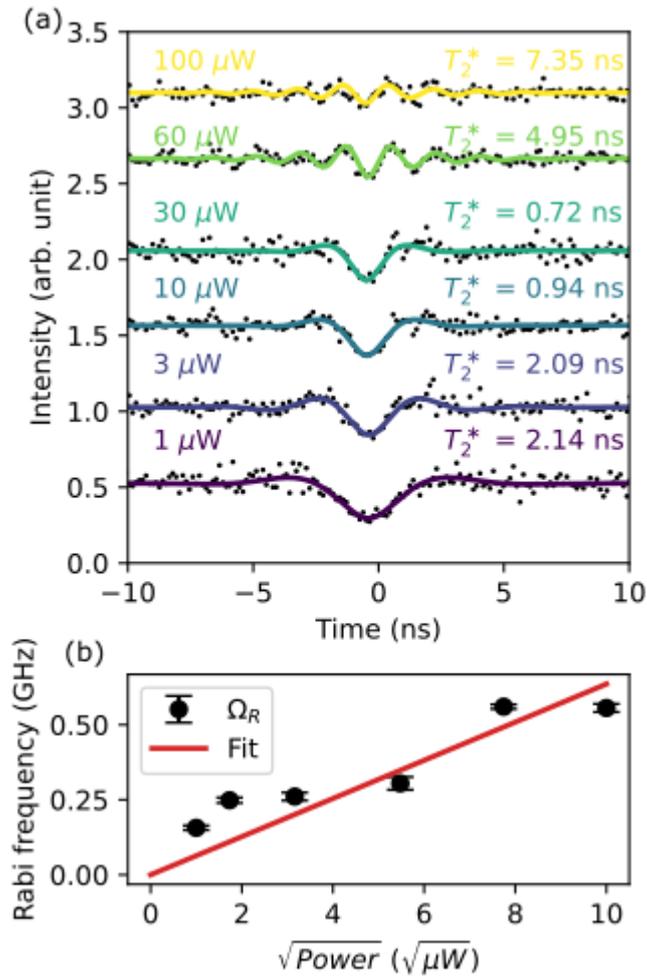

Fig S4. Rabi oscillations. a) Second order correlation data collected at the same site as in Fig. S3. Each data set is offset vertically for clarity, and fit with the sinusoidal exponential function given in text. b). The Rabi frequency extracted from the fit in (a) is plotted against the square root of the excitation power, and a linear function is used to fit the data.